\documentclass[final]{siamltex}
\usepackage{epsfig,graphicx}
\usepackage{epic,eepic}
\usepackage{float}
\title{Cut Size Statistics of Graph Bisection Heuristics\thanks{Division
  de Physique Th\'eorique, Institut de Physique Nucl\'eaire, Universit\'e\
Paris-Sud, F--91406 Orsay Cedex, France, Unit\'e de Recherche
  des Universit\'es Paris XI et Paris VI associ\'ee au C.N.R.S.}}

\author{
 G.R. Schreiber
 \thanks{also at: Service de Physique Th\'eorique, Orme des Merisiers,
 C.E.A.-Saclay, F - 91191 Gif-sur-Yvette Cedex, France,
 e-mail: {\tt georg@spht.saclay.cea.fr},
 {\tt http://ipnweb.in2p3.fr/\~{}schreibe} }
 \and
 O.C. Martin
 \thanks{
 e-mail: {\tt martino@ipno.in2p3.fr}},
 {\tt http://ipnweb.in2p3.fr/\~{}martino} }

\date{}

\begin{document}

\maketitle

\begin{abstract}
  We investigate the statistical properties of cut sizes generated
  by heuristic algorithms
  which solve approximately the graph bisection problem.
  On an ensemble of sparse random graphs, we find empirically that
  the distribution of the cut sizes found by ``local''
  algorithms becomes peaked as the number of vertices in the
  graphs becomes large. Evidence is given that
  this distribution tends towards a Gaussian whose mean and variance
  scales linearly with the number of vertices of the graphs.
  Given the distribution of cut sizes associated with each heuristic,
  we provide a ranking procedure which takes into account both the
  quality of the solutions and the speed of the algorithms.
  This procedure is demonstrated for a selection of local
  graph bisection heuristics.

\bigskip
\bigskip
\hspace{4cm}
Submitted to SIAM Journal on Optimization
\end{abstract}

\bigskip
\bigskip



\pagestyle{myheadings}
\thispagestyle{plain}
\markboth{ {\sc G.R. Schreiber and  O.C. Martin}}{{\sc Cut Size Statistics of
Graph Bisection Heuristics}}

\clearpage

\section{Introduction}
\label{section:introduction}

  Algorithms for tackling
  combinatorial optimization problems \cite{Papadimitriou_Steiglitz_82}
  may be divided into two classes. Exact algorithms such as
  exhaustive search, branch-and-bound, or branch-and-cut,
  form the first class; they determine (exactly) the optimum of the
  cost function which is to be minimized.
  However, for {\it NP}-hard
  problems, they require large computation ressources, and in
  particular, large computation times.
  The second class consists of ``heuristic'' algorithms; these
  are not guaranteed to find the optimal (lowest cost) solution, nor even
  a solution very close to the optimum, but
  in practice they find good approximate solutions very fast.
  For problems in science, one's main interest is in
  the optimal solution, so an exact algorithm is required.
  However, for many engineering applications, the heuristic
  approach may be preferable. There are several reasons for this:
  (i) The computational ressources are simply insufficient to solve
  the instances of interest by exact methods;
  (ii) The cost function
  one wants to minimize is
  computationally very demanding, and limited resources force one
  to use an approximate cost function instead. This is the rule
  rather than the exception with very complex systems such as VLSI.
  If the true cost function cannot be used, there is little
  point in finding the true optimum for the wrong problem.
  (iii) Heuristic algorithms typically generate numerous
  ``good enough'' solutions, thus providing
  information about the statistical properties of low cost
  solutions.
  This information can in turn be used for generating
  better heuristics, or for finding new criteria
  for guiding the branching in exact algorithms such as
  branch-and-bound.

  For almost any combinatorial optimization problem,
  it is very easy to devise heuristic algorithms which
  perform quite well; this is probably why so many such
  algorithms have been proposed to date.
  Usually they fall into just a few families, the most popular
  of which are
  local search, simulated annealing, tabu search, and
  evolutionary computation.
  The practitioner is frequently confronted with the problem of
  choosing which method to use.
  Thus he would like to rank these algorithms and
  determine which one is best for his
  ``instance'' (the set of parameters which completely specify the
  cost function).
  A difficulty then arises because most heuristic
  algorithms are stochastic, so that they can give many
  different solutions for a single instance. In general,
  the distributions of solution costs generated by
  the different heuristics overlap, so that the winning algorithm
  varies from one trial to another.
  Furthermore, it is necessary to balance the quality of the solutions
  found against the time necessary to find them since in practice
  heuristics run at very different speeds.
  The final goal of this paper is to do just this kind of balancing: in
  Section \ref{section:ranking} we shall
  introduce a generally applicable ranking method
  which is based on the possibility of performing
  multiple runs from random starts for each algorithm until
  an allotted amount of computer time is exhausted.
  Our ranking method then determines
  whether it is better to have a fast heuristic which gives not so good
  solutions or a slower heuristic which can give better solutions.

\par
  Establishing a ranking on a {\it single} instance may be
  what is needed for a real world problem, but it is not a
  useful prediction tool. It is preferable to consider the
  effectiveness of a heuristic when it is applied to a
  {\it family} of instances.
  Since a detailed knowledge of the distribution of costs
  is necessary for our ranking procedure, the major
  part of this paper is an in depth study of the
  {\it statistics} of costs found by
  several classes of heuristics.
  The NP-hard \cite{Garey_Johnson_79} combinatorial
  optimization problem chosen for our study is the graph
  bisection problem, hereafter simply called the graph
  partitioning problem (GPP).
  This choice is justified by the wide range of practical
  applications of the GPP. These
  include host scheduling \cite{Berger_Bokhari_87},
  memory paging and
  program segmentation \cite{Kernighan_Thesis},
  load balancing \cite{Leland_Hendrickson_94}, and
  numerous aspects of VLSI-design
  such as logic partitioning \cite{Johannes}
  and placement \cite{Dunlop_Kernighan_85,Kirkpatrick_Gelatt_Vecchi_83}.
  Because of these applications, the GPP has been used as a
  testing ground for many heuristics.
  For our work, a selection had to be made; in view of the previous
  studies by
  Johnson {\it et al.} \cite{Johnson_Aragon_McGeoch_Schevon_89},
  Lang and Rao \cite{LangRao_93}, and Berry and
  Goldberg \cite{BerryGoldberg_97},
  we have restricted our study
  to iterative improvement heuristics based on local search and to
  simulated annealing.
  Having made a choice of optimization problem and algorithms,
  it remains to define
  the class of instances for the testbeds.
  Ideally, this family of instances should reflect the structure
  of the actual instances of interest to the practitioner.
  Since we do not have a particular
  application in mind, we shall follow the studies of
  \cite{Johnson_Aragon_McGeoch_Schevon_89,LangRao_93,BerryGoldberg_97},
  and consider an ensemble of sparse random graphs.
  From our numerical study, we have found
  that all of the heuristics tested share
  the following properties when the random graphs become large:
  (i) each algorithm can be characterized by a fixed percentage
  excess above the optimum cost;
  (ii) the partitions generated have a {\it distribution} of costs
  which becomes peaked, both within a given graph and across all graphs;
  (iii) these distributions tend towards Gaussians.
  Because of these properties, our ranking of heuristics
  on large graphs is largely determined by
  the mean and variance of the costs found, and thus a constant
  speed-up factor has only a very small effect on the ranking.
  We expect this property to hold for most problems and
  heuristics of practical interest, leading to a very robust
  ranking.

  The paper is organized as follows. In Section \ref{section:mincut}
  we define the GPP as well as the ensemble of random graphs
  used for our testbed. Section \ref{section:randomcuts}
  derives properties of random partitions, and
  shows that the distribution of cut sizes has a
  relative width which goes to zero as the instance size grows.
  In Section \ref{section:statphys} we argue why
  this property should hold also for
  the distribution of costs found by {\it heuristic} algorithms
  based on local iterative processes.
  In Section \ref{section:algorithms} we discuss the
  heuristic algorithms we have included in our tests.
  Section \ref{section:selfaveraging} gives the mean
  and standard deviation of the costs found
  as a function of graph size; the distribution for the costs
  is indeed found to be peaked. This leads to
  a first ranking which, however, does not take into account
  computation times. To implement our speed-dependent
  ranking, we must determine the
  {\it distribution} of cut sizes found by the different algorithms.
  This is the subject of Section \ref{section:distribution}, where
  evidence is given that the distribution on any typical graph
  tends towards a Gaussian in the limit of large graphs.
  In Section \ref{section:ranking} we
  present our ranking method which takes into
  account both the quality of the solutions as well as the
  speed of the heuristics.
  In Section \ref{section:discussion}, finally,
  we discuss the results and conclude.

\section{Minimum cuts}
\label{section:mincut}

  The graph partitioning (or graph ``bisection'') problem
  (GPP) can be defined as follows. Consider a graph
  $G = (V, E)$
  which consists of a set of $N$ vertices
  $V = \{v_1, v_2, \ldots , v_N \}$
  and a set of (non-oriented) edges connecting pairs of vertices.
  It is convenient to introduce the matrix $E_{ij}$, called
  the connectivity matrix, given by

  \[E_{ij} = \left\{ \begin{array}{r@{\quad:\quad}l}
  1 & \mbox{ if } v_i \mbox{ is connected to } v_j \\
  0      & \mbox{otherwise}
  \end{array}\right.\]

\noindent
  Since the edges are
  non-oriented, $E_{ij} = E_{ji}$. (Some of what will
  be discussed applies to weighted graphs; then
  $E_{ij}$ will represent the weight of the $ij$ edge.)
  A partition of $G$ is
  given by dividing the vertices of $G$ into two disjoint subsets
  $V_1$ and $V_2$ such that
  $V = V_1 \cup V_2$. The number of edges connecting
  $V_1$ to $V_2$ is called the cut of the partition, and will
  be denoted by ${\cal C}$. It
  is given by

  \begin{equation}
  \label{costfunction}
  {\cal C}[V_1, V_2] = \sum_{i\in V_1,j\in V_2} E_{ij}.
  \end{equation}

\noindent
  The GPP (or ``Min-cut'' problem) consists of finding
  the partition $(V_1,V_2)$ for
  which the cost (\ref{costfunction}) is minimum subject to
  given constraints on the sizes of $V_1$ and $V_2$.
  The GPP is {\it NP}-hard \cite{Garey_Johnson_79}.
  In the standard formulation to which we shall restrict ourselves
  in this work,
  $V_1$ and $V_2$ have equal sizes.

  For our study, it is necessary to fix an ensemble of graphs
  for the testbed. We have chosen $G(N,p)$ the ensemble
  of random graphs of $N$ vertices where each edge is present
  with probability $p$. The choice of $G(N,p)$ is justified
  by its tractable mathematical properties and
  by the fact that many
  workers \cite{Johnson_Aragon_McGeoch_Schevon_89,LangRao_93,BerryGoldberg_97}
  have used graphs in this ensemble to test heuristics.
  The problem of finding the properties
  of the minimum cut size when the graphs belong to such
  an ensemble is sometimes called the
  {\it stochastic} GPP. Let us review some of the known results for
  this problem; this will serve to motivate our conjectures for
  the behavior of cuts obtained from {\it heuristics}.
  For each graph $G_i$, call ${\cal C}_0$ its minimum cut size.
  Taking $G_i$ from the ensemble $G(N,p)$,
  ${\cal C}_0$ is a random variable.
  Following derivations now standard
  in a number of other stochastic combinatorial optimization
  problems (COP), it is possible to show
  using Azuma's inequality \cite{Azuma_TMJ} that the distribution
  of ${\cal C}_0$ becomes peaked as $N \to \infty$. This means
  that as $N$ becomes large,
  $({\cal C}_0 - {\Big \langle} {\cal C}_0 {\Big \rangle} ) /
  {\Big \langle} {\cal C}_0 {\Big \rangle}$,
  the relative fluctuations about the mean,
  tend to zero. This property, often referred to as
  ``self-averaging'', is typical of processes to which many
  terms contribute. For certain stochastic COP, it is possible
  to show further that the mean minimum cost satisfies a
  power scaling law in $N$,
  so that ${\cal C}_0 / N^{\gamma}$ converges
  in probability to a limiting value as $N \to \infty$.
  In the case of the stochastic GPP, there is no proof
  that such property hold. Nevertheless, it is believed that
  such a scaling holds: within the $G(N,p)$ ensemble at $p$ fixed,
  calculations show that
  ${\cal C}_0 / N^2 \to p/4$ with probability one as
  $N \to \infty$ \cite{Fu_Anderson_86}.
  As will be shown in the next section,
  this is also the limiting behavior of random cuts, and so
  the ensemble at $p$ fixed is not a challenging one
  for heuristics. The reason for this ``uninteresting'' scaling
  is the high number of edges connecting
  to any vertex. Thus we consider in this work the ensemble
  $G(N,p)$, $p=\alpha / (N-1)$ with $\alpha$ fixed; $\alpha$ is
  the mean connectivity (number of
  neighbors of a vertex) of the graphs.
  These graphs are sparse, in contrast to the
  dense graphs obtained by taking $p$ to be independent of $N$.
  Consider the optimal partition. At a typical vertex in $V_1$,
  some finite fraction of its edges will connect to vertices
  in $V_2$. With each vertex contributing an $O(1)$ amount to the
  cut size, ${\cal C}_0$ is expected to grow linearly with $N$.
  Since ${\cal C}_0 / N$ is known to be peaked at large $N$, it is
  natural to conjecture the stronger property that
  ${\cal C}_0 / N$ tends towards a constant
  with probability one as
  $N \to \infty$.
  A major motivation for this work is our expectation that
  an identical scaling law should hold
  if we replace ${\cal C}_0$ by the cost
  found by a heuristic algorithm,
  albeit that the limiting constant depends on
  the heuristic. To motivate such a property, the next
  section analyzes the cut sizes of random partitions; then in
  Section \ref{section:statphys} we consider
  the ``statistical physics'' of the GPP so as
  to interpolate between the case of minimum cuts and that of
  random cuts.

\section{Cuts of random partitions}
\label{section:randomcuts}

  Here we show explicitly that a large $N$ scaling law
  holds for the cut sizes of random partitions, and that asymptotically
  these random cuts have a Gaussian distribution with a relative variance
  proportional to $1/N$.

\par
  Consider any graph in $G(N,p)$.
  One can always write the cut size of a random partition
  as ${\cal C} = X + Y$ where
  $X$ is the mean (random) cut size for the graph under consideration, and
  $\langle Y \rangle = 0$. ($\langle ~ \rangle$
  is the average over the random partitions.)
  Averaging explicitly over all balanced partitions of the fixed
  graph, we find
  $X = \sum E_{ij} N / [2 (N-1)]$.
  The interpretation of this formula is very simple: any edge of weight
  $E_{ij}$ has a probability $N / [2 (N-1)]$ of being cut.

  In the ensemble $G(N,p)$ of random graphs,
  it is easy to calculate the first few moments of $X$.
  In particular, we find
  ${\Big \langle} X {\Big \rangle}  = p N^2 / 4$ and
  ${\Big \langle} (X - {\Big \langle} X {\Big \rangle})^2 {\Big \rangle} =
  p (1-p) N^3 / [8 (N-1)]$.  (${\Big \langle} ~ {\Big \rangle}$ denotes
  the average over the ensemble $G(N,p)$.)
  We also see that $X$ is
  the sum of $M = N (N-1)/2$ {\it independent} random variables; this
  implies that the $k$th cumulant (connected moment) of the distribution
  of $X$ statisfies

  \begin{equation}
  {\Big \langle} X^k {\Big \rangle}_c = (N-1)^2  [{N \over {2 (N-1)}}] ^{k+1}
  {\Big \langle} E_{ij}^k {\Big \rangle}_c .
  \end{equation}

\noindent
  At large $N$, we then have
  ${\Big \langle} X^k {\Big \rangle}_c \sim N^2$ in
  the constant $p$ ensemble, and
  ${\Big \langle} X^k {\Big \rangle}_c \sim \alpha N$ in
  the $p \sim \alpha /N$ ensemble.

  The random variable $Y$ is more subtle as it is
  the sum of $M$ {\it correlated} variables.
  Nevertheless, for any graph, it is possible
  to compute the moments of $Y$, and we have done this
  explicitly for the second and third
  moments. (The expressions are too long
  to be given here.) If we average $Y^2$ both over random partitions
  and over $G(N,p)$, we obtain:

  \begin{equation}
  {\Big \langle} \langle Y^2 \rangle {\Big \rangle} =
  {p(1-p) \over 8} N^2 (N-2)/(N-1) .
  \end{equation}

\noindent
  The calculations get significantly more complicated for the
  higher moments. In order to keep to simple expressions, we
  limit ourselves to the ensemble with
  $p = \alpha / (N-1)$. Then we find:

  \begin{equation}
  {\Big \langle} \langle Y^2 \rangle {\Big \rangle} =
  {\alpha \over 8} N + O(1) , \qquad
  {\Big \langle} \langle Y^3 \rangle {\Big \rangle}
  = - {\alpha \over 8} + O({1 \over N}) .
  \end{equation}

\noindent
  Furthermore, the
  graph to graph fluctuations of $\langle Y^2 \rangle$ become
  negligible in relative magnitude, so that the ratio of a
  typical variance to the mean variance goes to $1$ at
  large $N$. This however is not true for the higher moments;
  for instance, we find that the typical value
  of $\langle Y^3 \rangle$ grows as $N^{1/2}$, but taking in
  addition the mean over graphs leads to a
  $N$ independent behavior. Finally, one can show that
  $\langle Y^k \rangle_c / \langle Y^2 \rangle^{k/2} \to 0$
  with probability one.
  This shows that as $N \to \infty$, $Y$ has a Gaussian distribution,
  of zero mean, and of variance growing linearly with $N$, whose
  coefficient is graph independent.

  Coming back to ${\cal C} = X + Y$, the cut size of a random partition,
  we find that the normalized correlation coefficients between
  powers of $X$ and $Y$ tend to zero
  at large $N$, and thus $X$ and $Y$ become independent random
  variables in that limit.
  This, along with the results previously derived,
  shows that at large $N$, ${\cal C}$ itself
  has a Gaussian distribution. From these results,
  we deduce the large $N$ behavior:

  \begin{equation}
  { {\Big \langle} \langle ({\cal C} - {\Big \langle \langle
  {\cal C} \rangle {\Big \rangle} )^2 \rangle {\Big \rangle} }
  \over {{\Big \langle} \langle {\cal C} \rangle {\Big \rangle}^2}  } \sim
  {4 \over {\alpha N}} ~,
  \end{equation}

\noindent
  so that {\it relative} deviations from
  the mean go to zero.
  Thus the distribution of $\cal C$ becomes peaked, and
  ${\cal C} / N  \to \alpha /4$ with
  probability one as $N \to \infty$.
  The convergence of the distribution of ${\cal C} / N$ to a
  ``delta'' function is referred to as the self-averaging of $\cal C$.

  The scaling of the variances
  can be summarized at large $N$ by writing
  \begin{equation} \label{eq:randcut}
  c \equiv { {\cal C} \over N} \sim
  {\Big \langle} \langle c \rangle {\Big \rangle}
  + \frac{\sigma_X^*}{\sqrt{N}} x
  + \frac{\sigma_Y^*}{\sqrt{N}} y
  \end{equation}

\noindent
  where $x$ and $y$ are independent Gaussian random variables
  of zero mean and unit variance; $\sigma_X^*=\sqrt{\alpha / 8}$ is the
  standard deviation (rescaled by $1/\sqrt{N}$)
  of $X$, and
  $\sigma_Y^*=\sqrt{\alpha / 8}$ that of $Y$.
  Thus $\sigma_Y^*$ describes the fluctuations of the cut sizes within
  a graph, and
  $\sigma_X^*$ describes the fluctuations of the mean cut size
  from graph to graph.

  We have used these analytical results to test the validity of our
  computer programs. The first two moments of $X$ allowed us
  to test our generation of random graphs in $G(N,p)$.
  Similarly, a check on our random number generator was obtained by
  verifying on several graphs
  that the second moment of $Y$
  found by the numerics was in agreement with our formulae.
  Finally, we also checked that random
  cut sizes have a limiting Gaussian distribution, with
  a third moment which scales to zero at large $N$.
  (For this check, we
  performed random partitions on $100 \, 000$ graphs for
  $N=100, 500, 1000$, and $2000$.)

\section{Statistical physics of the GPP}
\label{section:statphys}


  We saw that cut sizes of random partitions in $G(N,p)$ have
  a self-averaging property; we conjectured that this property
  also holds
  for the minimum cut. It is possible to interpolate between
  these two kinds of partitions (random and min-cut)
  by following the formalizm of
  statistical physics. For any given graph, consider
  the ``Boltzmann'' probability distribution $p_B$, defined for
  an arbitrary partition $P$ of cut size ${\cal C}(P)$:

  \begin{equation} \label{equation:boltzmann}
  p_B(P) = {e^{ -{\cal C}(P) / T }  \over Z } ~ .
  \end{equation}

\noindent
  $Z$ is chosen so that $p_B$ is normalized (a probability
  distribution) and $T$ is an arbitrary positive parameter called
  the temperature. When $T \to \infty$, we recover the ensemble
  of random partitions where all partitions are equally probable,
  while when $T \to 0$, the ensemble
  reduces to the partitions of
  minimum cut size. For intermediate values of the temperature,
  the partitions are weighted
  according to an exponential of their cut size. In
  this ``Boltzmann'' ensemble,
  one can define the moments of the cut sizes just as was done
  in the case of random partitions.
  In most statistical physics problems, it is possible to show that
  the quantity in the exponential of Eq. (\ref{equation:boltzmann})
  (here, the cut size) is self-averaging.
  For {\it random} graphs, however,
  the proofs are inapplicable; nevertheless,
  other evidence indicates that the cut size is self-averaging
  at any temperature \cite{Mezard_Parisi_Virasoro_87}.
  This self-averaging can be understood qualitatively at low temperature as
  follows. The number ${\cal N}({\cal C})$ of partitions of
  cut size $\cal C$ is a sharply increasing function of
  $\cal C$, whereas the Boltzmann factor is a sharply decreasing function
  of $\cal C$. Note that the probability distribution $P({\cal C})$ of
  $\cal C$ is given by the product of these two functions.
  Using naive but standard statistical physics arguments for
  ${\cal N}({\cal C})$, one finds that $P({\cal C})$ has a peak at
  ${\cal C}^*(T)$ which grows linearly with $N$ and that the width
  of the distribution is $O(\sqrt{N})$, which gives the
  self-averaging property for $\cal C$.
  In addition, this kind of argument says that $P({\cal C})$ becomes
  Gaussian at large $N$, a result which is usually correct
  in statistical physics systems.

  A number of statistical physics results have been obtained
  for the GPP in the
  ensemble of dense random graphs, {\it i.e.}, for $G(N,p)$ at $p$ fixed.
  In particular, highly technical
  calculations \cite{Mezard_Parisi_Virasoro_87,Fu_Anderson_86}
  indicate that the cut sizes are self-averaging at all temperatures,
  that is as $N \to \infty$, relative fluctuations within a fixed graph
  become negligible, as well as those from graph to graph.
  The mean cut size is given by

  \begin{equation}
  {\Big \langle} \langle {\cal C} \rangle {\Big \rangle} =
  {{p N^2} \over 4} - U(T) \sqrt{p(1-p)}N^{3/2} + O(N)
  \end{equation}

\noindent
  as $N \to \infty$.
  (If the mean over graphs is not performed, the formula
  remains valid for ``almost all'' sequences of
  graphs with $N \to \infty$.)
  In this equation, $U(T)$ is a function
  of temperature only, there is no dependence on
  $p$ as long as $p$ is independent of $N$.
  The limit $T \to 0$ gives the expected (and typical) value of the
  minimum cut, with
  $U(T=0)=0.3816.$. Although there is no proof yet that these
  calculations are exact, there is general agreement in the statistical
  physics community that the results are correct.

  The case of sparse random graphs ($p \sim 1 / N$)
  has also been studied within the statistical physics
  approach \cite{Banavar_Sherrington_Sourlas_87,de_Dominicis_Goldschmidt_89}.
  So far, however, the problem
  has proven to be intractable with no plausible
  solution in sight. Nevertheless, it is expected that
  the cut sizes are self-averaging at any temperature
  and that the mean of the
  distribution scales linearly with $N$ at large $N$.

  The property of self-averaging seems quite generic.
  The reason it should hold in these systems is that the
  cut size of a partition is the sum of a large number of
  random variables which are not {\it too} correlated.
  It is very plausible that the cut size is self-averaging
  whenever partitions are
  generated by an iterative process involving just a few vertices
  at a time.
  All local search methods, and modifications thereof
  such as simulated annealing, fall into this category.
  Thus our claim is
  that any heuristic algorithm
  which generates partitions iteratively according to
  local (in vertex space) criteria will lead to
  cut sizes which are self-averaging.
  Thus the distribution of cut sizes found by any such
  heuristic should become peaked as $N \to \infty$. Furthermore,
  in this limit,
  the distribution should converge towards a Gaussian in the way given
  by the central limit theorem.
  We will see in the sections to follow that this
  is indeed born out empirically for all of the heuristics which we
  have investigated.

  The arguments we have presented are not specific to the graph
  partitioning problem, so we expect them to apply to most stochastic COPs
  having many variables in their cost function.
  Surprisingly, there has been very little research
  on this topic.
  In the context of the ``NK'' model with binary variables,
  a study by Kauffman and Levin~\cite{KauffmanLevin87} found that the
  costs of {\it local minima} became peaked towards the value of a
  {\it random} cost as $N$ grew. (This peculiar property is due to the
  structure of the energy landscape in that model.)
  However, concerning the behavior of {\it heuristic}
  solutions, research has almost exclusively
  focused on the case of the
  Euclidean traveling salesman problem where points
  are laid out on the plane. Most practitioners in that field know
  that local search heuristics give rise to costs whose
  relative variance decreases as the number of points
  increases. Furthermore, it was observed by
  Johnson and McGeoch~\cite{JohnsonMcGeoch96} among others that
  the costs tend towards a fixed percentage excess above the optimum.
  Our purpose here is to show {\it how} this convergence occurs, albeit
  in a different combinatorial optimization problem, and to provide
  a theoretical framework for understanding where this behavior comes
  from. Also, we pay special attention to the distinction between
  fluctuations within an instance and from one instance
  to another. We believe
  our findings are quite general, and in particular that the
  ensemble of instances considered
  need not be based on points in a physical space.

\section{Algorithms used in the testbed}
\label{section:algorithms}

  In view of the previous arguments, we have restricted ourselves
  to local heuristics. Without trying to be complete nor
  representative, we have studied the statistics of cut sizes
  for three types of local search and four versions of
  simulated annealing algorithms. In this section we
  sketch the workings of these heuristics.
  In Sections \ref{section:selfaveraging} and \ref{section:distribution},
  we show that the same self-averaging properties
  hold for all these
  algorithms in spite of their significant differences. There is thus
  no reason to believe that our claims are affected by the
  details of such algorithms; rather, the properties are most likely
  generic to dynamics which are local.

\subsection{Kernighan-Lin (KL)}

  In simple local search, one performs elementary transformations
  to a feasible solution of the COP as long as they decrease the cost,
  a procedure sometimes called $\lambda$-opting \cite{Lin_65}.
  A more sophisticated version consists in using ``variable
  depth'' search: one builds a sequence of $p$ elementary
  transformations, usually according to a greedy criterion.
  $p$ is not set ahead of time, and depends on the sequence of
  costs found. The elementary transformations are not imposed to
  decrease the cost, but the sequence of length $p$ must do so if it
  is to be applied to the current solution.
  Such a procedure was first proposed by
  Kernighan and Lin \cite{Kernighan_Lin_70}, in fact in the
  framework of the GPP.  Hereafter
  we will refer to their algorithm as ``KL''.
  The elementary transformation they use is the exchange of a pair of
  vertices, one vertex in $V_1$ being exchanged for one in $V_2$.
  A sequence of such exchanges is built up in a greedy and tabu fashion
  by performing a ``sweep'' of all the vertices:
  at each step of the sweep, one finds the best (largest cost gain)
  pair to exchange
  among those vertices which have not yet been moved in the sweep
  (tabu condition).
  The sweep has length $N/2$. When the sweep is
  finished one finds the position $p$ along
  the sequence of exchanges generated
  where the cut size is minimum. If this minimum leads to an
  improved partition, the transformation of $p$ exchanges
  is performed on the partition and another sweep is initiated;
  otherwise the
  search is stopped and the partition is ``KL-opt'', {\it i.e.},
  it is a local minimum under KL.

  The KL algorithm is deterministic although it is possible to
  introduce stochasticity to break degeneracies
  in selecting the best pair to exchange.
  Its computational complexity is not easy to estimate because the number
  of sweeps is not known in advance. (This is a generic difficulty
  in estimating the speed of iterative improvement heuristics.)
  However, in practice, one finds that KL
  finishes in a ``small'' number of sweeps. Thus the computational complexity
  is estimated to be a few times that of performing the last sweep, known
  as the check-out sweep.
  For our study, we have used our
  own implementation of KL \cite{Martin_Otto_95}, which uses
  heaps to find the best pair to exchange at each step.
  For sparse graphs, this leads to
  ${\cal O}(N \ln N)$ operations per sweep.
  A nearly identical KL is provided
  in the Chaco software package, which
  gives sensibly identical results.
  A faster implementation of the algorithm has
  been given by Fiduccia and Mattheyses \cite{Fiduccia_Mattheyses_82}
  whenever the use of a radix sort is possible; then the
  time for each sweep is ${\cal O}(N)$.

  In terms of quality of solutions found, KL is quite good.  What
  is surprising is that although Kernighan and Lin
  proposed their method over $20$ years ago,
  KL remains relatively unchallenged,
  at least as a general purpose method applicable to any
  kind of graph, regardless of its structure.
  Of course, for special kinds of graphs, such as meshes,
  other heuristics (e.g., spectral bisection) perform
  better
  \cite{BerryGoldberg_97,Hendrickson_Leland_95b,
  Johnson_Aragon_McGeoch_Schevon_89,KarypisKumar_97}.

\subsection{A multilevel KL-algorithm: CHACO}
  The Chaco software package includes a number of heuristics
  for partitioning graphs.
  (For information about this package, see the {\em Chaco user's guide}
  \cite{Chaco_Manual_V2}.)
  For our purposes, we have used only its
  ``multilevel'' generalization of KL, hereafter
  referred to simply as CHACO. The CHACO algorithm is based on
  a coarse graining or ``compactification'' of the graph to be partitioned.
  At each level, vertices are paired using a matching
  algorithm, and paired vertices are then considered as the vertices
  of the next higher level of compactification. Because of this process,
  it is necessary to have weighted edges; the weights
  are also propagated to the higher level.  The compactification is
  repeated until a sufficiently small graph is obtained to which
  spectral bisection is
  applied to get a first partition. Then this partition is
  used as the starting
  partition in KL for the graph at the level below it. This process
  is recursive, until one obtains a KL-opt partition of the original
  graph. (Note that this construction is deterministic, and does not
  require an initial ``random'' partition.)
  Such a multilevel strategy has been very successful
  for unstructured $2$ and $3$ dimensional
  meshes \cite{Hendrickson_Leland_95b,KarypisKumar_97},
  both in terms of
  solution quality (much better than for KL alone),
  and in terms of speed (much faster than KL because of the
  hierarchical nature).
  However, the
  usefulness of CHACO on random graphs is not {\it a priori} obvious,
  both in terms of speed and quality of solutions.

\subsection{Simulated Annealing algorithms}
  We have choosen as a third comparative algorithm simulated
  annealing (SA). SA is based on a set of elementary moves, just
  like local search, but now moves which increase the cost
  are accepted with (low) probability. Because of this,
  it is sometimes appropriate to consider SA as
  a noisy local search method.
  Simulated annealing is really a family
  of algorithms. To include some of
  the different bells and whistles proposed for this algorithm,
  we have considered four variations.
  These are: (i) the SA as first introduced by
  Kirkpatrick {\it et al.} \cite{Kirkpatrick_Gelatt_Vecchi_83}
  (referred to as FSA) where
  the initial and final temperatures
  are fixed ahead of time by the user and where a predetermined
  number of trial moves are performed at each temperature;
  (ii) Kirkpatrick {\it et al.} also proposed to determine
  the initial and final temperatures of the schedule dynamically.
  They set the initial temperature
  at the beginning of the run using the criterion that
  about $80 \%$ of the trial moves are accepted
  at that temperature.
  Similarly, they stop the cooling if for 5
  cooling steps the energy does not decrease.
  We will refer to this method as KSA.
  (iii) Johnson {\it et al.} \cite{Johnson_Aragon_McGeoch_Schevon_89}
  improved the speed of this algorithm by allowing an early
  exit to the next temperature of the schedule; the condition they
  proposed for exiting is having accepted a minimum number of moves.
  Also they modified the termination criterion to having
  an acceptance rate less than a threshold value.
  We will refer to this version as JSA.
  All three of these SA methods use an exponential cooling schedule
  with a cooling factor of $0.95$.
  (iv) The last SA variation consists in using
  an {\it adaptive schedule} whereby the next temperature value is
  determined on the fly according to the energy fluctuations at the current
  temperature. We have choosen for this variation
  the implementation of van Laarhoven and Aarts
  \cite{van_Laarhoven_Aarts_85,van_Laarhoven_Aarts_87}.
  To obtain good results one would have
  to spend a long time in the ``freezing'' phase of the cooling. Since this
  would increase the computation times significantly we have choosen
  not to use a fine-tuned adaptive schedule but one which provides
  a cooling factor of the same magnitude as in the other SA algorithms
  presented. This allows us to have similar
  computation times for all the simulated annealing algorithms investigated.

  In SA, one can use the
  same elementary moves as in local search, {\it i.e.}, for
  the GPP, pair exchanges.  However, once a low cost
  partition is obtained, it will take a long time (or a lot of luck)
  to find further good exchanges. Finding a good {\it pair} is best done
  by finding the first vertex to transfer and then
  the second, {\it i.e.}, by using a {\it sequential} process. This suggests
  relaxing the constraint of having balanced partitions, and replacing
  it by a penalty function which keeps the
  sizes of $V_1$ and $V_2$ nearly equal (small {\it off-balance}).
  We have followed a slightly different approach where
  each move destroying the balance must be
  followed by a move restoring the balance.
  Then the Markov chain explores the partitions which are
  balanced and those with ``off-balance'' of $\pm 1$.
  It is easy to see that this method is equivalent to having the
  cost of all the other partitions equal infinity; at fixed
  temperature and for long chains, one generates partitions with cut sizes
  given by the Boltzmann factor, within the constraint for the
  ``off-balance''.
  Indeed, the succession of accept/reject
  decisions makes the global probability distribution
  Boltzmannian in this enlarged space, so that we guarantee the same
  convergence properties as in the standard case.

  Some remarks concerning our implementations are in order.
  First, at fixed temperature, we perform a certain number of
  ``sweeps''. In each sweep, every vertex is sequentially
  considered as a candidate
  for changing sides of the partition; if the move were to violate
  our limit on the ``off-balance'', the move is rejected
  (in fact, it simply is not considered). A sweep thus requires $O(N)$
  operations. Our sweeps use
  random {\it permutations} rather than a fixed
  or random ordering of the vertices. The use of
  random permutations should
  -- according to certain authors
  \cite{Johnson_Aragon_McGeoch_Schevon_89,
  van_Laarhoven_Aarts_85,van_Laarhoven_Aarts_87} --
  result in a enhancement of the quality of the solutions found.
  Second, the maximum number of sweeps at any temperature is set to
  $\alpha \lambda$, with
  $\lambda = 10$ for all of our implementations.
  For FSA and KSA, this is in fact the (actual) number of sweeps,
  so that their computational complexity is $O(\alpha \lambda N)$
  times the number of temperature steps used.
  The cases of JSA and ASA are more difficult to evaluate.
  In practice we find that JSA is faster than KSA, but not by more than
  a constant factor. ASA on the other hand spends quite a lot of time
  at intermediate temperatures, all the more so that $N$ increases;
  empirically, we have found an $O(N^{3/2})$ complexity.

  In terms of quality, we are aware of no systematic study
  on sparse random graphs. In a previous SA work on the GPP,
  Van Laarhoven and Aarts used an adaptive
  decrement rule \cite{van_Laarhoven_Aarts_85,van_Laarhoven_Aarts_87}
  and claim a gain of about $13\%$ over
  simpler non-adaptive algorithms. They also compared their results
  to those from
  the algorithm used by Johnson {\it et al.} for the GPP, who claimed
  an enhancement of about $5 \%$ for JSA
  over the Kernighan-Lin algorithm.
  The small gain found by Johnson \cite{Johnson_Aragon_McGeoch_Schevon_89}
  is, according to van Laarhoven and Aarts
  \cite{van_Laarhoven_Aarts_85,van_Laarhoven_Aarts_87},
  due to the use of a non-adaptive choice of the temperature decrement
  rule. However, we have found for sparse random graphs that the
  different variants of simulated annealing
  are nearly indistinguishable in terms of quality of solutions.
  This may be due to our not using a penalty term
  or to the different nature of the graphs used in the present study.

\subsection{Chained-Local-Optimization (CLO)}
\label{subsec:CLOalgorithm}

  The chained-local-optimization (CLO) strategy is a synthesis of
  local search and of simulated annealing \cite{Martin_Otto_96}.
  The essential idea is to have simulated annealing
  sample not all solutions, but only locally optimal solutions.
  This strategy
  is guaranteed to be at least as good as local search, and
  has been successfully applied to the
  traveling salesman problem \cite{Martin_Otto_Felten_92} and
  to the partitioning of unstructured meshes \cite{Martin_Otto_95}.

  In this work, we use KL as the local search engine.
  Given any initial KL-opt partition $P_i$, the simplest implementation
  of CLO will: (i) apply
  a perturbation or ``kick'' to modify significantly the
  partition (in practice
  this means exchanging {\it clusters} of vertices); (ii)
  run KL on the modified partition so as to reach a new
  KL-opt partition$P_f$; (iii) apply the accept/reject procedure
  for going from the initial partition ($P_i$) to the final one ($P_f$).
  This defines the analogue of one move of a simulated
  annealing algorithm, except that many modifications to the partition
  have occured in this single step. The temperature may be modified
  according to a schedule if desired, but for simplicity,
  we have set the temperature to zero in all of our runs.

  As was discussed in the context of simulated annealing,
  it is inefficient to exchange vertices or clusters simultaneously,
  it is better to do it sequentially.
  Our present CLO algorithm thus proceeds as
  follows. Given $P_i$ an initial balanced KL-opt partition,
  choose a (connected) cluster of $p$ vertices in $V_1$ (or $V_2$), and
  move them into $V_2$ (respectively $V_1$). KL-optimize
  this partition to generate
  an intermediate (off-balanced) partition.
  Now choose a cluster of $p$ vertices in $V_2$ ($V_1$) and move them into
  $V_1$ ($V_2$); KL-optimize this modified partition to generate $P_f$, the
  final (and {\it balanced}) partition. This whole procedure is our
  ``simulated annealing'' step, and we apply the accept/reject
  criterion for going from $P_i$ to $P_f$.

  When runing CLO on irregular meshes \cite{Martin_Otto_95},
  it was possible to perform large kicks, exchanging many vertices
  at once. Unfortunately, for sparse random graphs, we find
  that the acceptance when doing so becomes low. We have thus used
  ``small'' kicks, creating clusters of sizes varying randomly between
  $3$ and $13$. Given such small kicks, KL usually terminates in just $2$
  sweeps, and the speed of CLO per kick is about half that of KL.

  Consider now the limit of large $N$. Using the analogy with
  simulated annealing, if a fixed ($N$-independent) number
  of small kicks are used, it can be expected
  that CLO will perform no better
  than KL itself. We have thus chosen to use a number of kicks which
  scales linearly in $N$, namely $\lambda N$ with $\lambda = 0.1$.
  This choice of course influences the quality of the solutions
  generated, a larger value of $\lambda$ giving {\it a priori} better
  results. The computational complexity of this
  algorithm is then of order $N^2 \log (N)$.

\section{Self-averaging of the cut size}
\label{section:selfaveraging}

  In the rest of this paper, we study the
  statistical properties of the cut sizes generated by
  the algorithms described in Section \ref{section:algorithms} when
  applied to random initial partitions. The
  ensemble of graphs used is that of random graphs
  with mean connectivity $\alpha = p (N-1) = 5$ (see
  Section \ref{section:mincut}).  This value was chosen
  because at much larger connectivities, the ratio between the
  best and worst cut size approaches $1$, and at lower connectivities,
  algorithms taking explicit advantage of disconnected parts of the graph
  will outperform general purpose heuristics.
  In order to minimize effects associated with our finite sample of
  graphs in the ensemble, we have benchmarked all the algorithms on the
  {\it same} graphs. The number of
  graphs used during the production runs was $10\;000$ with values of $N$
  ranging between $50$ and $200$;
  however, because the
  CHACO algorithm was fast, we have also performed runs on
  $100\;000$ graphs for that heuristic.

  The purpose of this section is to give numerical
  evidence that the distribution of cut sizes
  becomes peaked in the limit of large graphs,
  for each of the heuristics considered.
  (Further properties of the distribution will be given in
  Section \ref{section:distribution}.)
  We find that each algorithm
  generates cut sizes for which both the mean and variance
  scale linearly in $N$.
  From this behavior, it is clear that
  the distribution of cut sizes becomes peaked at large $N$, {\it i.e.},
  that the cut sizes are self-averaging.
  Also, assuming (cf. Section \ref{section:mincut}) that the minimum
  ({\it i.e.}, optimum) cut size scales linearly with $N$ at large $N$,
  we then see that each
  heuristic algorithm leads to a fixed
  percentage excess above the true optimum.
  (Note that the worst cut size also has a linear scaling in $N$.)
  This percentage excess provides a first ranking
  of the algorithms, which, however, does not take into account
  the speed of execution.

  If ${\cal C}(i,m)$ is the cut obtained by a heuristic
  for the graph $G_i$ and an initial partition $m$,
  define the mean cut per vertex
  ${\Big \langle} \langle {c} \rangle {\Big \rangle}$ by:

  \begin{equation}
  {\Big \langle} \langle {c} \rangle {\Big \rangle} \equiv
  {\Big \langle} \langle {{\cal C}(i,m) \over N} \rangle {\Big \rangle}~,
  \end{equation}

\noindent
  where the averages are over initial partitions and over
  the ensemble of graphs studied (cf. Section 3 for the notation).
  We compute these ensemble averages numerically using
  the standard estimator (hereafter, overlines refer to numerical
  averages):

  \begin{equation}
  {\overline c} \equiv
  { \sum_i \sum_m {\cal C}(i,m) \over N \sum_i \sum_m 1} \approx
  {\Big \langle} \langle {c} \rangle {\Big \rangle} ~.
  \end{equation}

\noindent
  The approximation is due to a statistical error {\it e}
  associated with fluctuations of ${\cal C}(i,m)$ both with $m$ and $i$.
  It is not difficult to see that for our problem, one does not need to
  perform an average over $m$; using
  any finite number $R$ of partitions for
  each graph $G_i$ provides an unbiassed estimator
  of ${\Big \langle} \langle {c} \rangle {\Big \rangle}$.
  Furthermore, the statistical error {\it e} is not
  very sensitive to $R$,
  making it numerically inefficient to take a large value for $R$.
  Because of this, we have performed the numerical averages with $R=1$,
  and this leads to a simple expression for
  {\it e}, the statistical error on $\overline c$:

  \begin{equation}
  {\it e}^2 =
  { {\Big \langle} \langle ({c} - \langle {\Big \langle} {c}
   \rangle {\Big \rangle} )^2 \rangle {\Big \rangle} \over
  {\sum_i 1} } \approx
  { ( {\overline {c^2}} - {\overline {c}}^2 )  \over {\sum_i 1} }.
  \end{equation}

\begin{figure}[htb]
  \begin{center}
  \hspace*{-1.0cm}\includegraphics[scale=0.40]{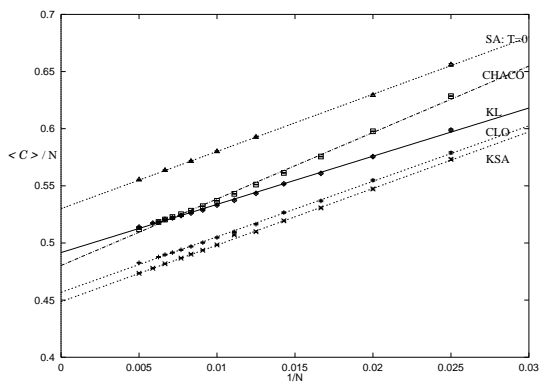}
  \caption{Scaled mean cut sizes for the different algorithms.}
  \label{fig:meanOfc}
  \bigskip
  \end{center}\end{figure}

\noindent
  Figure \ref{fig:meanOfc} shows the dependence of $\overline c$
  on $1/N$. (The error bars are
  too small to be visible. Also, in order to avoid cluttering the
  figure, we have included among the simulated annealing algorithms
  only KSA; the other implementations of simulated annealing
  give nearly identical results.)

\noindent
  For all algorithms, the figure suggests that there is
  a limiting large $N$ value
  for $\overline c$ and that the convergence to this limit is
  linear in $1/N$.
  We have thus fitted the data to a linear function:

  \[\frac{\overline{{\cal C}}}{N} \equiv \overline{c}
  \approx A + \frac{B}{N}.\]

\noindent
  The values of the $A$ and $B$ coefficients obtained from
  the fits are given in Table \ref{tab:scaling}, and
  the $\chi^2$ values show that the fits are good.

  \begin{table}[h]
  \begin{center}
  \parbox{8cm}{
  \begin{center}
  \begin{tabular}{|c|c|c|c|}\hline
  algorithm & $A$ & $B$ & $\%$ excess \\ \hline
  KSA    &  0.4485  &  4.95  &  0.00  \\
  FSA    &  0.4489  &  4.92  &  0.08  \\
  ASA    &  0.4499  &  4.96  &  0.32  \\
  JSA    &  0.4513  &  4.88  &  0.63  \\
  CLO    &  0.4568  &  4.85  &  1.8   \\
  CHACO  &  0.4802  &  5.81  &  7.1   \\
  KL     &  0.4916  &  4.21  &  9.6   \\
  SA $T=0$ & 0.5302  & 4.79  & 18.2   \\ \hline
  \end{tabular}
  \end{center}
  \caption{Estimates for the large $N$ value and slope of the
           mean cut size per vertex
           and percentage excess relative to the KSA heuristic.}
  \label{tab:scaling}}
  \end{center}
  \end{table}

\begin{figure}[h]
  \begin{center}
  \hspace*{-1.0cm}\includegraphics[scale=0.40]{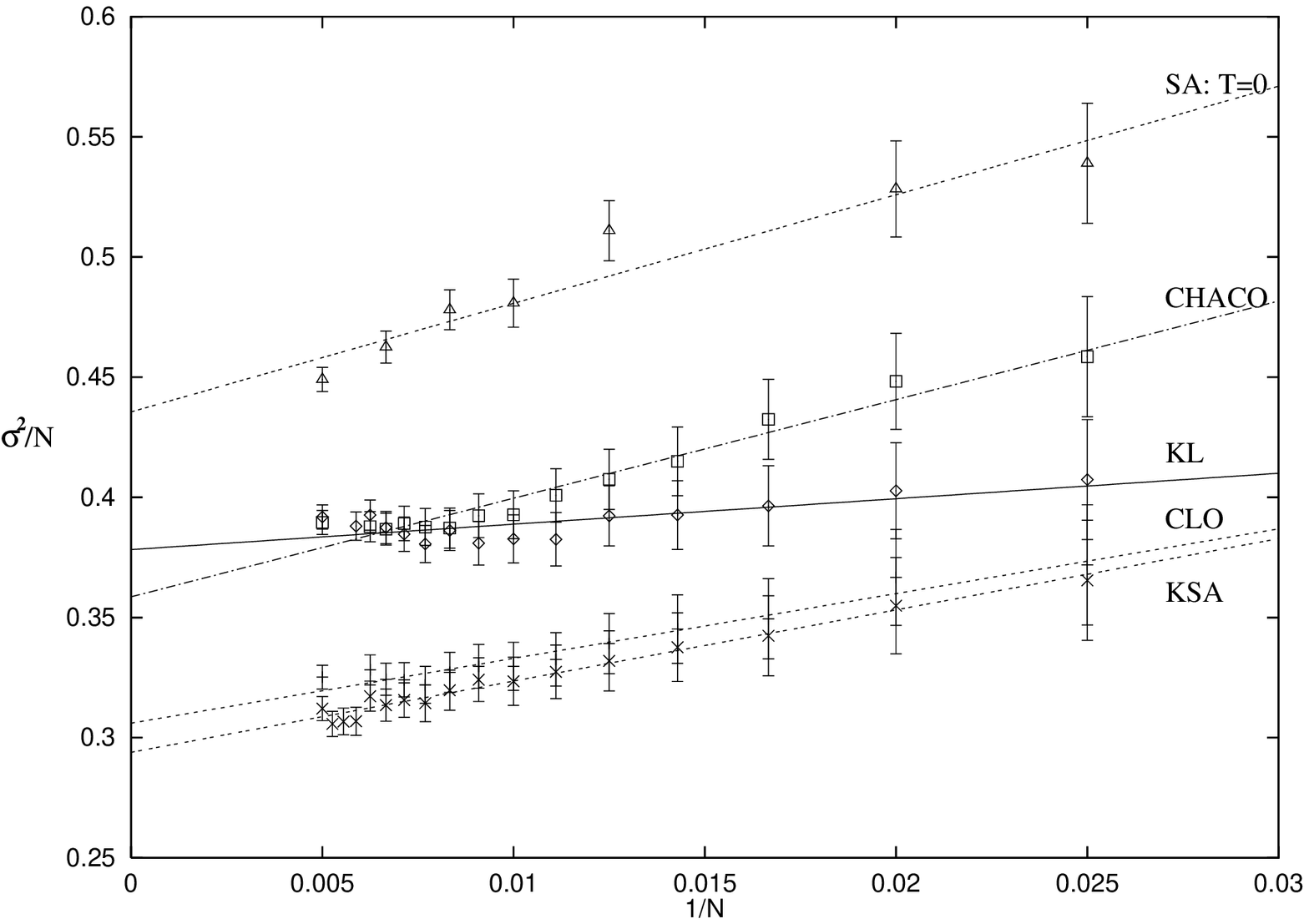}
  \caption{Scaled variance of the cut sizes for the different algorithms.}
  \label{fig:varianceOfc}
  \bigskip
  \end{center}\end{figure}

  An identical analysis can be performed on the {\it variance}
  of the cuts found by the different algorithms.
  Figure \ref{fig:varianceOfc} shows the dependence on $N$
  for the rescaled quantity
  $N ( {\overline {c^2}} - {\overline c}^2 )$. The
  scaling in $N$ is apparent, just as it was for $\overline c$.

  In summary, our data lead us to conclude that
  the mean and variance of $\cal C$
  scale linearly with $N$ at large $N$. Then the relative width of the
  distribution of $\cal C$ is proportional to $1 / \sqrt{N}$,
  showing that
  the distribution for the cut sizes becomes peaked for all
  the algorithms investigated. (One can also say that the distribution of
  ${\cal C}(i,m)/N$ tends towards a delta function as
  $N \to \infty$, which is what we mean by self-averaging.)
  Since the fluctuations of ${\cal C}(i,m)$
  include both graph to graph fluctuations and fluctuations within
  a graph, we can conclude that the relative fluctuations within
  a fixed typical graph necessarily also go to zero. (N.B.:
  although for our runs we use $R=1$, our observable
  $( {\overline {c^2}} - {\overline {c}}^2 )$
  is an unbiassed estimator for
  ${\Big \langle} \langle ({c} - \langle {\Big \langle} {c}
   \rangle {\Big \rangle} )^2 \rangle {\Big \rangle}$
  which includes both types of fluctuations.)
  Thus in the large $N$ limit,
  each algorithm will give a fixed percentage excess above
  the minimum for almost all graphs and almost all random initial partitions.

\subsubsection*{A speed independent ranking}

  Since each algorithm is characterized by a percentage
  excess, we can introduce a ranking of the different heuristics
  according to their excess in the
  large $N$ limit. (Of course, this ranking does not
  take into account the speed of the algorithms!)
  For our graphs and our
  implementation of the different heuristics, the winners
  are in the class of simulated annealing.
  The best is KSA; using this as the reference rather than the
  true min cut size (which is unknown),
  JSA has an excess of $0.63\%$,
  ASA an excess of $0.32\%$, and FSA an excess of $0.08\%$.
  The next best heuristic is
  the CLO-algorithm, followed by CHACO,
  and finally KL. (The results for the excesses are given
  in Table \ref{tab:scaling}.)
  We have also included for general
  interest the excess obtained by a
  zero temperature ``simulated annealing'': $18.21 \%$;
  note that it gives much less
  good results than KL, while true simulated annealing gives much better
  results than KL.

  As a comment, let us remark that the relative solution quality
  of the algorithms is determined to higher precision
  than the absolute quality. Simply put, the cut sizes
  we obtain for the different algorithms are correlated because
  they are performed
  on the same graphs, so that
  the statistical error on
  ${\Big \langle} \langle {c}_{CLO} - {c}_{KL} \rangle {\Big \rangle}$
  for instance is $3.2$ times smaller than the statistical error on
  ${\Big \langle} \langle {c}_{CLO} \rangle {\Big \rangle}$
  alone. This is why it is possible to give reliable values
  for the excesses of the different simulated annealing algorithms
  even though their solution quality is very similar.
  Nevertheless the ranking for the simulated annealing algorithms is
  not without ambiguity. The FSA algorithm is, for larger $N$, within
  the statistical error of the KSA algorithm, and hence we have no
  strong evidence that one is better than the other.

  The other algorithms are easily ranked. KL and CHACO
  are $9.6 \%$ and  $7.1 \%$ worse than KSA,
  but CLO is only $1.8 \%$ worse.
  The comparison with KL is qualitatively (though not quantitatively)
  similar to that given by
  Johnson {\em et al.} \cite{Johnson_Aragon_McGeoch_Schevon_89}
  and by van Laarhoven and Aarts
  \cite{van_Laarhoven_Aarts_87}.
  Both claimed a gain of the SA-algorithm over the
  KL-algorithm of about $5\%$ and $13\%$, respectively.
  The differences with our results
  have several origins. First, we have performed an average over an
  ensemble of graphs. Second, our graphs have
  slightly different characteristics from
  the ones they use. Third we have not introduced a
  penalty term in our implementation of
  simulated annealing; this probably affects the quality
  of the solutions found.

\section{Distribution of cut sizes}
\label{section:distribution}

  In this section we deepen our statistical study of $\cal C$.
  As shown in the previous section, the distribution of ${\cal C} / N$
  tends towards a delta function; it is natural to ask {\it how} this
  limit is reached, and to understand the nature of intra- and inter-graph
  fluctuations.  It is convenient to use the
  framework introduced in Section \ref{section:randomcuts} but where
  {\it random} partitions are replaced by the partitions found by applying
  one of our heuristics to a random start.
  For each graph $G_i$, and each initial partition $m$, we define

  \[ {\cal C}(i,m) = X(i) + Y(i,m)  \]

\noindent
  where $\langle Y(i,m) \rangle = 0$ so that $X(i)$ is the average cut size
  found on graph $G_i$, and $Y(i,m)$ gives the fluctuation of the
  cut size about its mean for that graph. For each of our heuristics,
  our study indicates that for a large random graph $G_i$,
  $Y$ has a nearly Gaussian distribution, and that the width of
  this distribution is essentially
  independent of $i$. We study this distribution
  at large $N$ and show that its width
  is self-averaging and that its relative asymmetry goes
  to zero. Finally, we have evidence that $X$ and $Y$ become
  independent variables at large $N$. These properties will lead to
  a fast and robust ranking of the heuristics in
  Section \ref{section:ranking}.

\begin{figure}[h]
  \begin{center}
  \hspace*{-1.0cm}\includegraphics[scale=0.40]{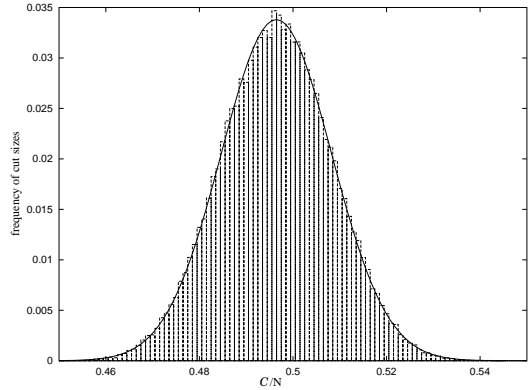}
  \caption{Histogram of KL cut sizes for one $N=1000$ graph with overlaid
           Gaussian.}
  \label{fig:gaussian}
  \bigskip
  \end{center}\end{figure}

  Figure \ref{fig:gaussian} shows the distribution of
  cut sizes found by KL on one
  $N=1000$ graph chosen at random from $G(N,p)$ with $p=\alpha/(N-1)$.
  Superposed
  is a Gaussian with the same mean and variance. The figure
  gives good evidence that the distribution of $Y$ for that graph
  is very close to a Gaussian. Then an obvious question is whether
  the distribution of $Y$ is similar across different graphs.
  For each of our heuristics, we find that the answer is yes,
  as indicated by the following study of the moments of $Y$.
  (Note that for the CHACO algorithm, the default parameter setting
  generates the initial starting partition
  deterministically by application of the coarse graining strategy,
  then a spectral method is applied. Since there is no ``random''
  initial partition,
  there are no fluctuations in the cut size as a function of $m$ and
  so little in this section applies to CHACO with these parameter settings.)

\begin{figure}[h]
  \begin{center}
  \hspace*{-1.0cm}\includegraphics[scale=0.40]{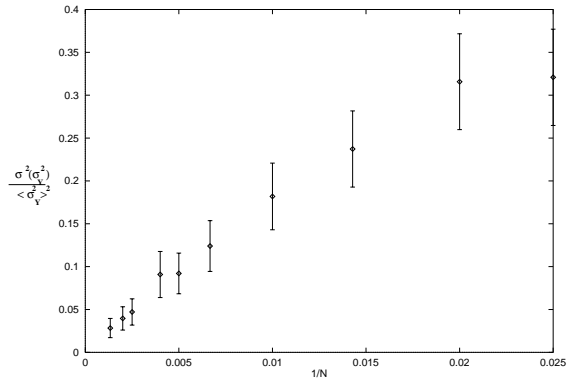}
  \caption{Relative variance of the intra-graph cut size variance
$\sigma^2_Y$.}
  \label{fig:selfaveragingofsigma2Y}
  \bigskip
  \end{center}\end{figure}

  To quantify how $\sigma^2_Y(i) \equiv \langle Y^2(i,m) \rangle$ varies
  from graph to graph, we measured its mean and variance over $i$.
  First, we measured the ensemble averages
  ${\Big \langle} \sigma^2_Y(i) {\Big \rangle}/N$.
  For each heuristic, the data extrapolates to a
  limiting value as $N$ becomes large.
  Comparing with the results for the mean cut size, we find
  that the algorithms which lead
  to the best cut sizes also have the smallest widths for the
  $Y$ distribution.
  Second, we studied the
  {\it variance} of $\sigma^2_Y(i)$,
  {\it i.e.}, $\sigma^2\left(\sigma^2_Y(i)\right)$.
  This study requires high statistics,
  and so was performed to high accuracy only for KL, the fastest
  of our algorithms; however the other algorithms show qualitatively
  the same behavior. Figure \ref{fig:selfaveragingofsigma2Y}
  displays for KL the $1/N$ dependence of
  the relative variance of $\sigma^2_Y(i)$, {\it i.e.},
  the inter-graph variance of $\sigma^2_Y(i)$ divided by the square
  of its mean. As can be seen from the figure, the ratio
  goes to zero at large $N$, showing that $\sigma^2_Y(i)$ is self-averaging.
  Simply put, this means that the width (over $m$) of the $Y$ distribution
  has relative fluctuations from graph to graph which dissapear
  as $N \to \infty$. (Our lower statistics data for the other heuristics
  are consistent with this conclusion.)

  Following the statistical physics analogy given
  in Section \ref{section:statphys},
  there is reason to believe that the distribution
  of $Y$ tends towards a Gaussian as in the case of random partitions.
  To test this conjecture, we have
  measured the asymmetry of the distribution of $Y$ on numerous graphs
  for KL.  First, we find that the typical asymmetry is small, and that the
  mean of the third moment of $Y$ satisfies

  \[ {\Big \langle} \langle Y^3(i,m) \rangle {\Big \rangle} /
  {\Big \langle} \sigma^2_Y(i) {\Big \rangle}^{3/2}  \to 0 \]

\noindent
  as $N \to \infty$. Second, we have
  checked that the average of the squared asymmetry is also small, {\it i.e.},

  \[ {\Big \langle} \langle Y^3(i,m) \rangle ^2{\Big \rangle} /
  {\Big \langle} \sigma^2_Y(i) {\Big \rangle}^{3}  \to 0 . \]

\noindent
  These properties give strong evidence that the distribution of $Y$
  for any graph tends towards a Gaussian of zero mean and of variance
  $A N$ as
  $N \to \infty$, where $A$ depends on the heurisitic but not on
  the actual graph.

  The distribution of $X(i)$ can be studied similarly. The previous
  section gave its mean as a function of $N$ and also
  showed that it is self-averaging. It is of interest
  to quantify the decrease with $N$ of its relative variance.
  We have found that the distribution of $X$
  is roughly compatible with a Gaussian distribution of width
  proportional to $\sqrt{N}$
  for each of the algorithms. (Unfortunately, a quantitive test
  of this requires very high statistics.)
  However, the distribution of $X(i)$ is
  not essential for our ranking procedure as will be clear in the next
  section, so we have not studied it in greater depth.

  Finally, to completely specify the statistics of ${\cal C}(i,m)$,
  it is necessary to describe the correlations between $X(i)$
  and $Y(i,m)$. We have found numerically
  that these variables are nearly uncorrelated, with in particular
  the correlation between $X(i)$ and $\sigma^2_Y(i)$ tending towards
  zero as $N \to \infty$. Assuming that this holds and that
  $X$ has a Gaussian distribution, then the distribution
  of ${\cal C}(i,m)$ is also Gaussian. Our measurement of the
  asymetry (jointly over $i$ and $m$) of ${\cal C}(i,m)$
  is compatible with this property at
  large $N$. (The total variance is then given by the sum of the
  variances of $X$ and $Y$.) This can be summarized mathematically
  by introducing two Gaussian random variables $x$ and $y$ of zero mean
  and unit variance, and modeling the rescaled cut size as the following sum:

  \[c(i,m) \sim {\Big \langle} \langle c \rangle {\Big \rangle} +
  \frac{\sigma_X^*}{\sqrt{N}} ~ x(i)
  + \frac{\sigma_Y^*}{\sqrt{N}} ~ y(i,m).\]

\noindent
  This equation is then the exact analogue of what was derived
  for the cut sizes of random partitions (see Eq. \ref{eq:randcut}).

\section{A speed dependent ranking of heuristics}
\label{section:ranking}

  In this section we come back to the initial motivation for this
  work, namely the necessity of
  comparing heuristics of very different speeds.
  The possibility of doing so is very relevant, as for most
  combinatorial optimization problems local search is
  quite fast and simulated annealing notoriously slow.
  Any meaningful ranking must
  determine whether it is
  better to have a fast heuristic which gives not so good
  solutions, or a slower heuristic giving better solutions.
  We now show how to introduce such a ranking when considering first
  just one graph, and then generalize to an
  ensemble of graphs. Finally, we illustrate what this ranking gives
  in the case of the heuristics in our testbed when applied to
  sparse random graphs.

\subsection*{The case of one graph}
  Consider a single graph $G$ on which one is to
  provide a ranking of
  a number of heuristics which give various
  cut sizes and run at different speeds. To take
  into account both the speed of the algorithms and the quality
  of the solutions they generate, we
  fix the amount of computation time allotted per algorithm. Call
  this time $\tau$ (measured for instance in CPU seconds on a given machine).
  Each heuristic then generates (non-optimal) solutions during
  that time using multiple random initial starts. Suppose that the
  speed of the algorithm of interest is such that $k$ independent
  starts can be performed in the allotted time $\tau$.  (We shall assume
  that the execution
  time is insensitive to the random initial start, as this is
  the case in practice with our heuristics.
  Knowledge of the speed of the algorithm then
  gives the value of $k$ which can be used.)
  For each start, there is an output or
  ``best-found'' cost.
  The output at the end of the $k$ starts is
  the best of these $k$ costs, hereafter called
  ``best-of-k''. The different algorithms are then
  ranked on the basis of the {\it ensemble mean} of their ``best-of-k''
  (the value of $k$ depending on $\tau$ and on the algorithm).
  This ensemble average is the average over the random
  numbers used both for the random initial starts and
  for running the algorithms (if any). This establishes a
  ranking for a particular graph
  and for a given amount of computation time $\tau$.

  It is inefficient to perform the average just mentioned in a
  ``direct'' way, {\it i.e.}, by extracting values of
  ``best-of-k'' over many multiple runs; it is far
  better to compute the average
  starting with the {\it distribution} of the ``best-found''
  cut sizes associated with single
  random starts. Call $P({\cal C})$ the probability of finding
  a ``best-found'' cut size of value ${\cal C}$,
  and $Q({\cal C})$ the associated cumulative
  distribution, {\it i.e.}, the probability of finding a cut size
  (strictly) smaller than ${\cal C}$.
  Since the cut sizes are integer valued, we then have
  $P({\cal C}) = Q({\cal C}+1) - Q({\cal C})$.
  Introducing the analogous probabilities ${\tilde P}_k$
  and ${\tilde Q}_k$ for
  the ``best-of-k'' values, one has:

  \[ 1 - {\tilde Q}_k({\cal C}) = (1 - Q({\cal C}) )^k . \]

\noindent
  The distribution
  for ``best-of-k'' can thus be generated from that
  of ``best-found'', and then ${\cal C}^*$, the mean of
  ``best-of-k'', is easily extracted.
  (This construction explains why we studied
  the distribution of single cut sizes
  in Section \ref{section:distribution}.)
  Note also that it is possible to extract ${\cal C}^*$ for a whole range of
  $\tau$ values with essentially no extra work
  since $\tau$ affects only $k$ and
  the determination of the mean of ``best-of-k'' represents a negligible
  amount of work once the distribution of ``best-found'' is known.

  The quantity ${\cal C}^*$ is in effect a
  quantitative measure of the effectiveness of the algorithm.
  Of course, ${\cal C}^*$ depends on the amount of
  computation ressources allotted,
  {\it i.e.}, $\tau$. As $\tau$ increases, $k$ increases (in jumps of unity),
  and ${\cal C}^*$ decreases. The broader the
  distribution of ``best-found'', the faster the decrease
  of ${\cal C}^*$ and the
  more useful it is to perform multiple runs.

  To establish the ranking, simply order
  the algorithms according to their
  ${\cal C}^*$. In general, this ranking may depend
  on $\tau$, and clearly it is sensitive to the lower tail of the
  distribution of ``best-found''.
  Let us illustrate this by considering
  for instance two heuristics $H_1$ and
  $H_2$ having two overlapping
  distributions for ``best-found'', with
  averages satisfying $\langle {\cal C}_{H_1} \rangle
                       <  \langle {\cal C}_{H_2} \rangle$.
  In the mean, $H_1$ seems better than $H_2$,
  but if $H_2$ is significantly faster,
  and if the tail of its distribution extends well into the domain
  of ${\cal C}_{H_1}$, then
  one can have ${\cal C}_{H_2}^* < {\cal C}_{H_1}^*$.
  $H_2$ may then be the more effective algorithm, assuming of
  course that $\tau$ is large enough
  so that indeed $H_2$ can be run multiple times.
  Some general properties may be derived
  assuming for instance that
  ${\cal C}_{H_1}$ and ${\cal C}_{H_2}$ are described
  by the same distribution but are
  shifted with
  respect to one another. Then if the tail of the distribution falls
  off as an exponential or faster,
  $H_2$ will {\it not} become more effective than $H_1$ as $\tau \to \infty$.

\subsection*{Ranking on an ensemble of graphs}
  The extension of this ranking to an ensemble of
  graphs is straight-forward.
  Assume that ${\cal C}^*$ is known
  for each graph $G$ and for each heuristic.
  ${\cal C}^*$ is a (real number) measure of the effectiveness of the
  heuristic on that graph, given an amount of computation time $\tau$.
  We can then generalize this measure from one graph to
  an ensemble of graphs by considering
  ${\Big \langle} {\cal C}^* {\Big \rangle}$, the
  mean of ${\cal C}^*$ over the relevant ensemble.
  The final ranking is then simply given by
  the ordering of the algorithms according to their mean effectiveness.

  Our expectation is that in a relatively
  homogeneous ensemble, the effectiveness (and thus the
  ranking) will be nearly the same for essentially
  all sufficiently large graphs and so the average behavior is also the
  typical behavior. We can expect this to happen whenever
  the distribution of cut sizes associated with the different
  heuristics do not overlap too much and have the same pattern
  regardless of the graph. This is what occurs in the case
  of our ensemble of random graphs: indeed, we saw that each algorithm
  leads to a fixed percentage excess cost at large $N$
  and that the distribution of costs is peaked. Then two algorithms
  have non overlapping distributions as $N \to \infty$ (unless they
  give rise to the same percentage excess). It is then clear that at large
  $N$, the mean ranking is the same as the typical ranking. It is
  also clear that
  increasing the amount of computer resources ($\tau$ and thus $k$)
  or speeding up an algorithm while keeping the
  quality of its solutions the same does very little to
  improve its ranking.

\subsection*{Illustration}
\begin{figure}[h]
  \begin{center}
  \hspace*{-1.0cm}\includegraphics[scale=0.60]{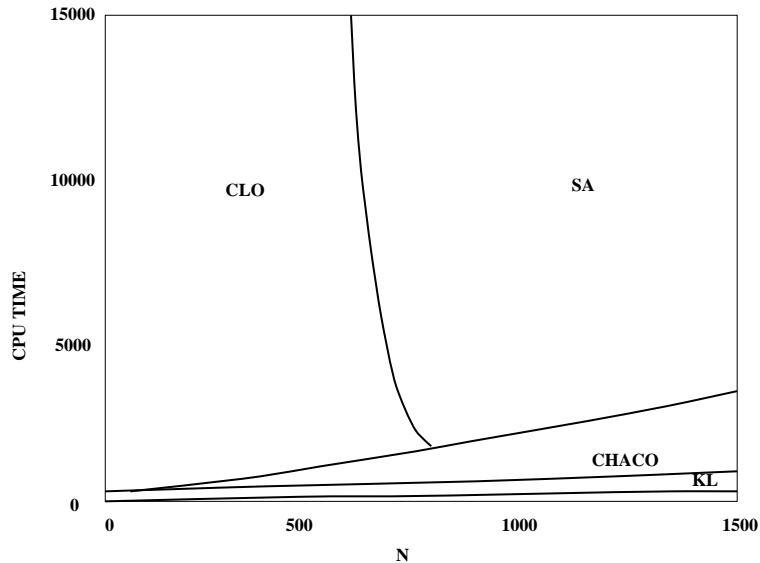}
  \caption{Ranking diagram}
  \label{fig:diagram}
  \bigskip
  \end{center}\end{figure}

  For each value of $N$ and $\tau$,
  we can follow the procedure just given to obtain ${\cal C}^*$
  for the different heuristics of interest for any given graph $G$, and repeat
  this for many graphs in $G(N,p)$.
  There are, however, a number of possible speed-ups in our case
  because of the statistical properties derived in the previous sections.
  First, although in principle the ``best-of-k'' construction
  has to be repeated for each graph,
  the results of Section \ref{section:distribution} provide
  a short-cut. Since the distribution for
  ``best-found'' is (to high accuracy) Gaussian,
  it is possible to map the mean of ``best-found'' to that of
  ``best-of-k'' once and for all:
  the mapping is just a shift by a $k$-dependent number of
  standard deviations.
  Second, noting that at fixed $N$, the variance of this Gaussian
  as well as the speed of the algorithm is essentially constant from
  graph to graph, we can
  calculate ${\Big \langle} {\cal C}^* {\Big \rangle}$ (the average
  over graphs) in terms of:
  (i) the CPU time necessary to find one ``best-found'';
  (ii) the mean cut size, ${\Big \langle} X(i) {\Big \rangle}$;
  (iii) the variance of the intra-graph cut
  sizes, $\langle Y^2(i,m) \rangle$, which is graph independent at large $N$.
  These quantities were measured for a number of values of $N$,
  and then fits were performed to interpolate to arbitrary values
  of $N$. From these fits, it is possible to
  compute analytically the values of
  ${\Big \langle} {\cal C}^* {\Big \rangle}$
  for any values of $N$ and $\tau$, and in particular the ``winning''
  algorithm (the first in our ranking).
  From this, define regions in ($N$,$\tau$) space
  where a given heuristic is the winner, leading to a ``diagram''
  as in Figure \ref{fig:diagram}.

  In our construction of this diagram, we have included JSA
  in our ranking but not FSA, KSA, nor ASA. This is because
  for our choice of parameters,
  all of the simulated annealing algorithms
  tested give very similar quality solutions,
  but JSA is slightly faster. Although the {\it effectiveness}
  of all these SA algorithms are nearly identical, their ranking depends
  on $N$ and $\tau$ because of the discrete jumps in $k$.
  (Whenever one algorithm increases its $k$ before the others, it may
  change its ranking.) In the diagram of Figure \ref{fig:diagram}, we have
  labeled the different regions according to the associated
  ``winner'', and have indicated
  the boundaries separating them. (Again, because of the discrete
  nature of $k$, we have smoothed these curves.)
  The labeling ``SA'' in fact corresponds to JSA.
  The CPU time is expressed in multiples of CPU-cycles. To give
  these units a machine independent and less technical meaning,
  it is enough to say that the lower boundary of the CHACO region
  corresponds to the time CHACO needs to run once.

  From this diagram, we see that at large $N$, given enough CPU time,
  the best algorithm is simulated annealing, simply because
  its mean excess cost is lower than that of the other algorithms.
  In this limit, the distributions for the cut sizes overlap very little,
  so the ranking is relatively
  insensitive to the algorithm's speed: using multiple
  random starts does {\it very} little to improve the quality
  of the solutions found as fluctuations about the mean become
  negligible. At smaller values of $N$, the fluctuations arising from
  different random starts are not negligible,
  so faster algorithms can outperform
  simulated annealing by using the best of $k$ runs. If we compare
  KL, CHACO, and CLO, we see that CLO is a bit slower but
  leads to substantially better solutions, and so is the winner
  if the amount of CPU time is enough for it to run. The other
  algorithms are competitive only if neither CLO nor simulated annealing
  can terminate a run.
  This explains why the KL region is nearly invisible, squeezed under
  the CHACO region, itself below the CLO and SA region.
  (Note: (i) on our random graphs, CHACO is {\it slower} than KL;
  (ii) the initial partition is set deterministically within
  the default settings of CHACO, so that its
  ``best-found'' and ``best-of-k'' values are identical.)

\section{Discussion and Conclusions}
\label{section:discussion}

  We have studied the {\it statistics} of cut sizes generated by
  graph partitioning heuristics,
  both within a given graph and over an ensemble of graphs. Motivated
  by a statistical physics analogy and by what happens for
  random partitions (Section \ref{section:randomcuts}), we
  obtained strong numerical evidence that the cut sizes
  generated on sparse random graphs
  are self-averaging, {\it i.e.}, that their
  distribution becomes peaked as the number of vertices $N$
  becomes large.
  (Quantitatively, this simply means that
  the {\it relative} fluctuations about the mean tend
  tend to zero as $N \to \infty$.) For the mean cut size,
  we found a linear dependence on $N$, indicating that
  each heuristic leads to a fixed percentage excess cut size above
  the true minimum. We expect analogous properties to hold for
  all local heuristics applied to any combinatorial optimization
  problem in which each variable is coupled to just a few others.

  We also investigated how the distribution of cut sizes approaches
  its limiting large $N$ behavior, and gave evidence that
  on typical graphs the distribution of cut sizes generated
  becomes Gaussian as $N \to \infty$. In that limit, each heuristic
  is then characterized by a mean cut size (over all graphs) and a
  variance describing the fluctuations in the cut sizes on any
  typical graph. This variance seems to scale linearly with $N$
  in the large $N$ limit and to be self-averaging also.

  The principal motivation for this work was to introduce a method to
  rank heuristics while taking into account both the quality of the
  solutions found and the speed of the algorithms.
  Knowledge of the distribution of cut sizes allows one to establish a
  meaningful ranking of the heuristics by assuming that the
  algorithms may be applied to $k$ different random starts,
  with the best of the $k$ runs giving the final cost.  Although this
  ranking can be done
  by brute force, we have used the properties just
  described to demonstrate it on the
  heuristics in our testbed. At ``large'' values of $N$ ($N > 700$),
  the winner is almost always simulated annealing. In fact,
  at large $N$, the distributions associated with the algorithms
  we have tested
  do not overlap significantly, so that the use of multiple
  runs to explore the tail of the distributions is not effective.
  For smaller values of $N$, the faster algorithms are more
  competitive, and we find that the winner is CLO except when the
  allotted time is too short for running even one run of CLO.
  Since the graph to graph fluctuations in the variance of the
  cut sizes found are small, this ranking ``in the mean'' is also
  in almost all cases the ranking on individual graphs; it is thus
  very robust.

  A number of questions remain open. How can one characterize the
  distribution of $X(i)$, the mean cut size on graph $i$?
  To what extent do similar properties hold for heuristics
  which are manifestly not local?
  Can the information found help generate better heuristics?
  Concerning this last question, it is worth pointing out
  that although simulated annealing is
  a general purpose method, it outperforms the other heuristics
  which were specifically developped for the graph partitioning
  problem. This suggests that some improvements in these methods
  might be obtainable by suitable modifications.

\section{Acknowledgement}
  We are indebted to Bruce Hendrickson and Robert Leland for providing
  us with their software package Chaco 2.0. We also thanks S. W. Otto
  and N. Sourlas for stimulating discussions.
  G.R.S. acknowledges support from an Individual
  EC research grant under contract number ERBCHBICT941665, and
  O.C.M. acknowledges support from the Institut Universitaire de
  France.
  Furthermore, G.R.S. would like to express his
  gratitude to Professor J.M. G\'omez G\'omez for his generous
  hospitality at the Department of Theoretical Physics of the
  {\em Universidad Complutense de Madrid},
  where part of this work was accomplished.



\bibliographystyle{siam}
\bibliography{../../bibinput/litbank,../../bibinput/co}

\end{document}